\documentclass{emulateapj}

\usepackage{lineno}
\usepackage{graphicx, subfigure}
\usepackage{amssymb}
\usepackage{amsmath}
\usepackage{epstopdf}
 \usepackage{hyperref}
\usepackage{color}
\usepackage{placeins}
\usepackage{natbib}
\usepackage{enumerate}
\usepackage[retainorgcmds]{IEEEtrantools}
\usepackage{textcomp}
\usepackage{footmisc}
\usepackage{xspace}
\usepackage[normalem]{ulem}

\newcommand{\Fermi}{\emph{Fermi}\xspace}

\newcommand{\hsi}{\emph{RHESSI}\xspace}
\newcommand{\goes}{\emph{GOES}\xspace}

\def\p0{$\pi^{\rm 0}$}

\def\de{$^{\circ}$\xspace}

\def\p8{\texttt{Pass8}}
\def\p7{\texttt{Pass7REP}}

\shorttitle{Sep 10 2017 Solar Flare}
\shortauthors{Omodei et al.}

\begin{document}

\title{\Fermi-LAT observations of the 2017 September 10$^{th}$ solar flare}

\author{Nicola~Omodei}
\email{nicola.omodei@stanford.edu}
\affiliation{W. W. Hansen Experimental Physics Laboratory, Kavli Institute for Particle Astrophysics and Cosmology, Department of Physics and SLAC National Accelerator Laboratory, Stanford University, Stanford, CA 94305, USA}

\author{Melissa~Pesce-Rollins}
\email{melissa.pesce.rollins@pi.infn.it}
\affiliation{Istituto Nazionale di Fisica Nucleare, Sezione di Pisa, I-56127 Pisa, Italy}

\author{Francesco~Longo}
\email{francesco.longo@trieste.infn.it}
\affiliation{Istituto Nazionale di Fisica Nucleare, Sezione di Trieste, I-34127 Trieste, Italy}
\affiliation{Dipartimento di Fisica, Universit\`a di Trieste, I-34127 Trieste, Italy}

\author{Alice~Allafort}
\affiliation{W. W. Hansen Experimental Physics Laboratory, Kavli Institute for Particle Astrophysics and Cosmology, Department of Physics and SLAC National Accelerator Laboratory, Stanford University, Stanford, CA 94305, USA}

\author{S{\"a}m~Krucker}
\affiliation{Space Science Laboratory, University of California, Berkeley, CA 94720-7450, USA}
\affiliation{University of Applied Sciences and Arts Northwestern Switzerland, CH-5210 Windisch, Switzerland}

\begin{abstract}
  The \Fermi-Large Area Telescope (LAT) detection of the X8.2 GOES class solar flare of 2017 September 10 provides for the first time observations of a long duration high-energy gamma-ray flare associated with a Ground Level Enhancement (GLE). The $>$100~MeV emission from this flare lasted for more than 12 hours covering both the impulsive and extended phase. We present the localization of the gamma-ray emission and find that it is consistent with the active region (AR) from which the flare occurred over a period lasting more than 6 hours contrary to what was found for the 2012 March 7 flares. The temporal variation of the proton index inferred from the gamma-ray data seems to suggest two phases in acceleration of the proton population. Based on timing arguments we interpret the second phase to be tied to the acceleration mechanism powering the GLE, believed to be particle acceleration at a coronal shock driven by the CME.

\end{abstract}


\section{Introduction}
High-energy gamma-ray solar flares provide the unique opportunity to examine pion-decay emission at the Sun. In order for this emission to occur, $>$300 MeV protons must be accelerated and subsequently interact with the chromosphere. Observations of prolonged pion-decay emission from flares~\citep{forr85,1993A&AS...97..349K,chup09,2000SSRv...93..581R} brought forth the idea that solar energetic particles (SEP) could be linked to these long-duration gamma-ray flares (LDGRFs) and that they could have a common origin, coronal Type II shocks~\citet{ram87}.

\Fermi-LAT~\citep{LATPaper} observations of the Sun have drastically increased the population of LDGRFs including hour-long emission from flares originating from active regions (AR) located behind the visible disk of the Sun~\cite{0004-637X-787-1-15,0004-637X-789-1-20,2017ApJ...835..219A}.  These observations suggest that the coronal mass ejection (CME) driven shock is the accelerating agent of the $>$300~MeV protons responsible for the $>$100~MeV emission. In fact, all of the LDGRFs observed by \Fermi-LAT are associated with fast CMEs. SEPs are also thought to be accelerated via shocks  and thus it is natural to search for a link between LDGRFs, SEP and CMEs (Winter et al. paper submitted to ApJ). Extreme and/or gradual SEP events often associated with ground level enhancements (GLE) are excellent test cases to investigate the connection with LDGRFs, but unfortunately the solar cycle 24 has been very poor in GLEs, with only 2 detected.

In this paper we present the \Fermi-LAT observations of 2017 September 10 solar flare associated with the second GLE of the solar cycle. We present time resolved localization of the $>$100~MeV emission and spectral evolution including the inferred proton index during the more 12 hour duration of emission.

\section{Observations and data analysis}
After almost an entire year of a nearly spotless Sun and no flaring activity, the largest flare
(GOES class X9.3) of the solar cycle erupted from active region 2673 on 2017 September 6. 
This flare was very bright in gamma-rays and the emission detected by the \Fermi-LAT lasted for almost 15 hours (see the Astronomers Telegram, ATel, 10720). 
Only 4 days later on September 10 at 15:35 UT a GOES X8.2 class flare (SOL2017-09-10) erupted from the same active region which had now moved to the edge of the western solar limb (S08W88). This flare lead to a gradual SEP event with proton energies measured by the GOES spacecraft exceeding 700 MeV/n and a very fast CME erupting over the western limb. 
The first appearance of the CME by LASCO C2 was at 16:00:07 UT and the initial speed was 3620 km s$^{-1}$. This flare was also associated with the second GLE (\#72) of this solar cycle. 
The GLE 72 onset was observed by several neutron monitors at 16:15 UT but the strongest increase in count rate was observed at 16:30 UT at the  Dome C station~(Mishev et al. submitted to Solar Physics).

\hsi has good coverage of SOL2017-09-10 with high resolution imaging showing a single hard X-ray footpoint located above the solar limb. The corresponding footpoint on the second flare ribbon appears to be occulted from Earth view. The visible hard X-ray footpoint is observed to be above the limb, co-spatial with the optical flare signal seen by SDO/HMI \citep[see][]{2015ApJ...802...19K}. The absence of STEREO B imaging of the flare ribbon location makes it impossible to precisely locate it relative to the limb. However, we firmly conclude that no hard X-ray emission is detected on the disk indicating that no part of the flare ribbons are on disk as seen from Earth view.
The $>$100~MeV emission detected by the LAT lasted for 12 hours and for that time period the Sun was the brightest gamma-ray source in the sky (see ATel 10721 for further details). The onset time for the LAT was found to be at 15:56 UT, the peak flux occurred at 15:59 UT remaining statistically significant until 05:11 UT  of September 11. During the flare, the LAT detected 130 photons with measured energy greater than 1~GeV and reconstructed direction less than 1$^{\circ}$ from the center of the solar disk. 
\begin{figure}[htb!]
\begin{center}
\includegraphics[trim=0.5cm 3cm 2cm 2cm, clip=true, width=0.48\textwidth]{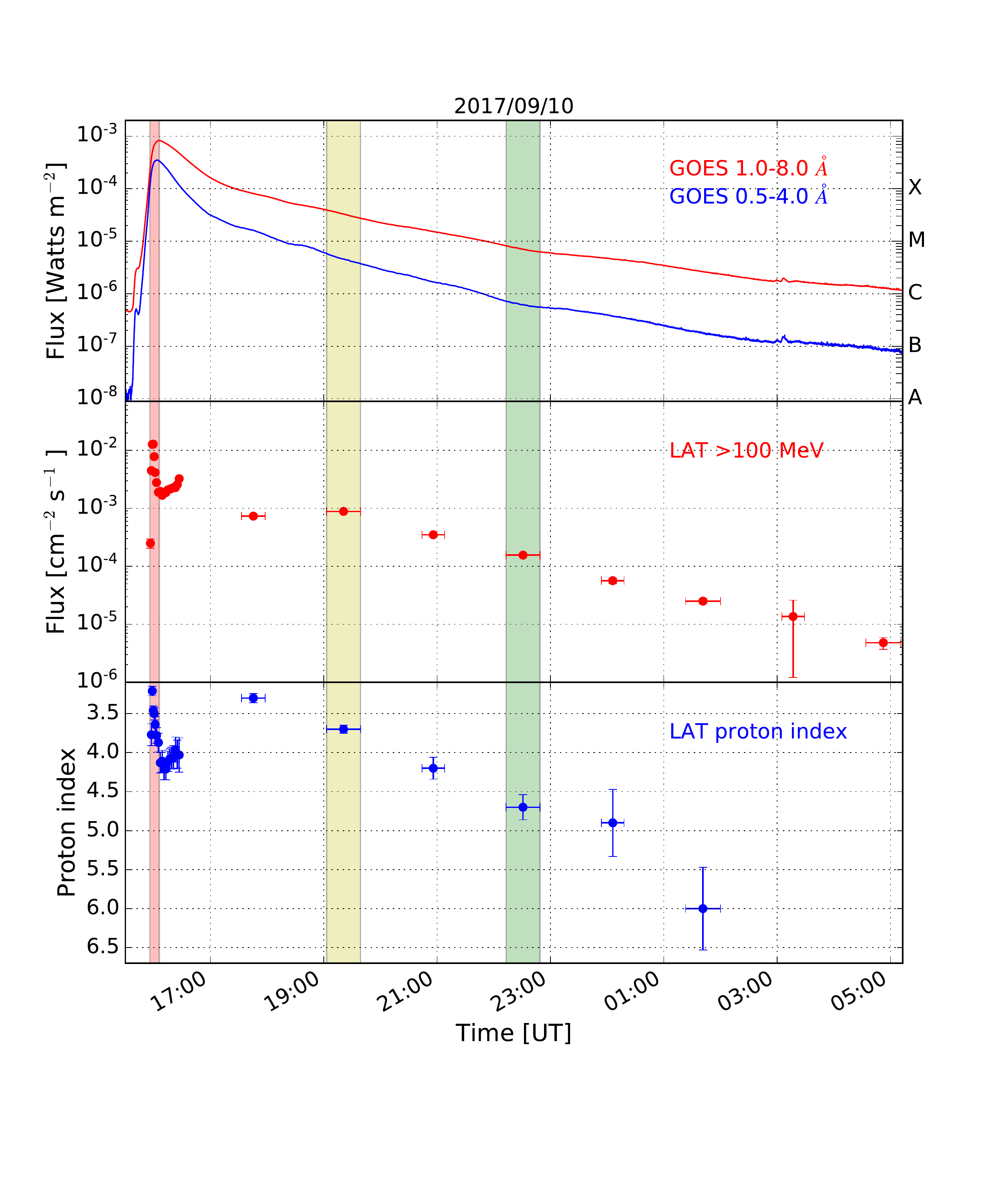}
\caption{Composite light curve for the 2017 September 10 flare with data from \goes X-rays, \Fermi-LAT $>$100 MeV flux and the best proton index inferred from the LAT gamma-ray data. The three color bands represent the time windows over which we performed the localization of the emission, shown in Figure~\ref{sept10_spatial_evolution}}.
\label{fig:SOL20170910_fullLC}
\end{center}
\end{figure}

\begin{figure}[htb!]
\begin{center}
\includegraphics[trim=0.5cm 3cm 2cm 2cm, clip=true, width=0.48\textwidth]{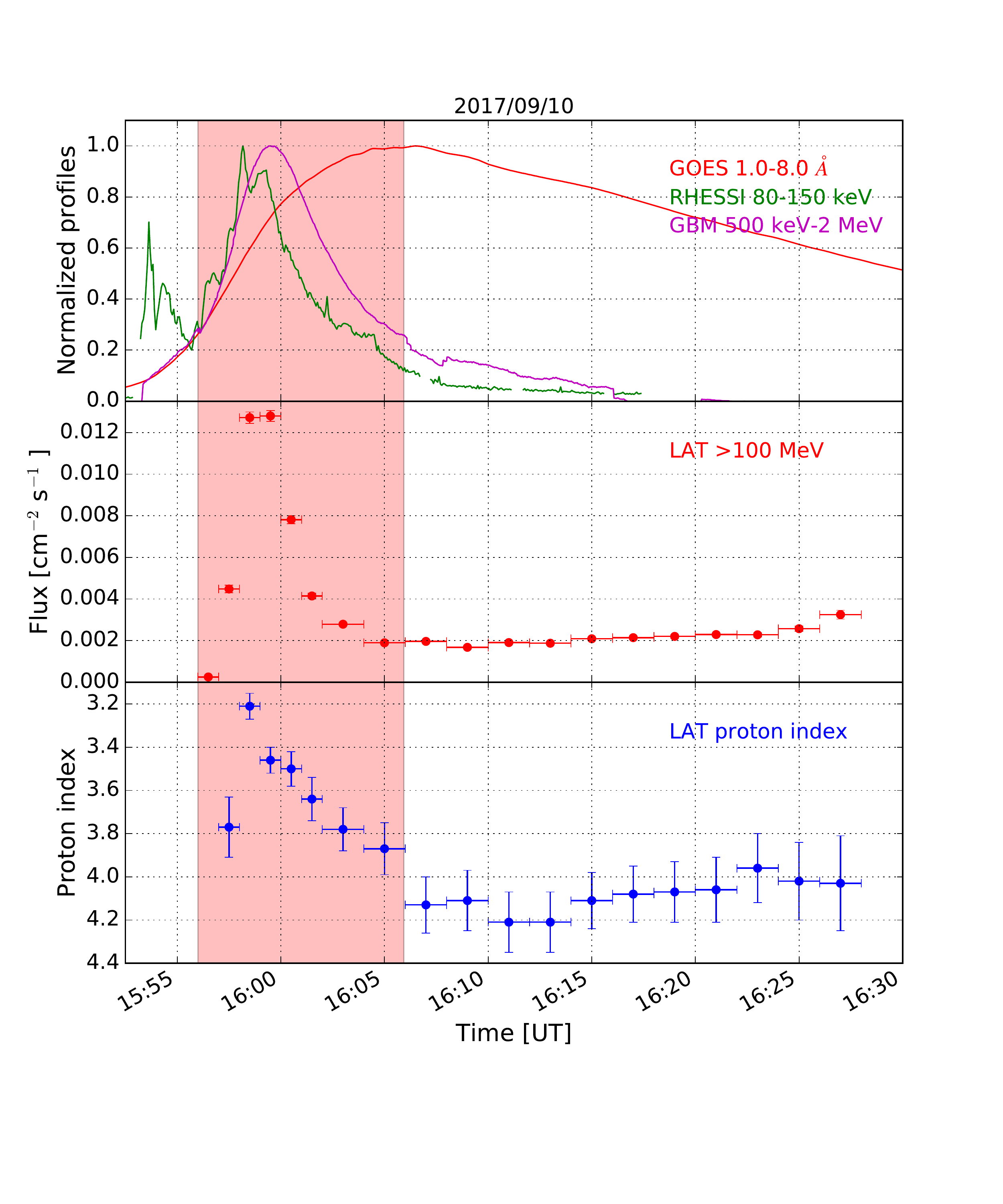}
\caption{Composite light curve for the impulsive phase of the 2017 September 10 flare with data from \goes, \hsi, \Fermi-GBM, and \Fermi-LAT. Bottom panel reports the best proton index in each time interval in which the LAT detected the flare. Only GBM-BGO data are shown because NaI suffered from pile-up. The red color band represents the time window over which the localization was performed and is shown in red in Figure~\ref{sept10_spatial_evolution}}.
\label{fig:SOL20170910_zoomLC}
\end{center}
\end{figure}


In Fig.~\ref{fig:SOL20170910_fullLC} we plot the light curves from \goes, and \Fermi-LAT for the full 12 hour detection period, while in Fig.~\ref{fig:SOL20170910_zoomLC} we plot \goes, \hsi, \Fermi-GBM, and \Fermi-LAT for the impulsive phase only. 
The bottom panel of each figure reports the best proton index in each time interval in which the LAT detected the flare. In section~\ref{sec:analysis} we describe how we obtain the protons index from the gamma-ray emission.  

\subsection{Spectral analysis}
\label{sec:analysis}
We performed an unbinned likelihood analysis of the \Fermi-LAT data with the \texttt{gtlike} program distributed with the \Fermi \texttt{ScienceTools}\footnote{We used the version 11-05-03 available from the \Fermi Science Support Center \url{http://fermi.gsfc.nasa.gov/ssc/}}. In order to avoid possible effects from pile-up in the anti-coincidence detector of the LAT during the brightest phase of the flare, from 15:54 to 16:28 UT, we selected the Pass 8 Solar flare Transient class (S15)\footnote{TRANSIENT015s available in the extended photon data through the Fermi Science Support Center} to perform our spectral analysis. This new transient class was developed to be insensitive to the high flux of X-rays often present during bright solar flares. For the remainder of the observation time (from 17:33 to the end of the detection), we used Pass 8 Source class events. For the entire detection time we used selected photons from a 10$^{\circ}$ circular region centered on the Sun and within 100\de from the local zenith (to reduce contamination from the Earth limb). 

We fit three models to the \Fermi-LAT gamma-ray spectral data. The first two, a pure power law (PL) and a power-law with an exponential cut-off (PLEXP) are phenomenological functions that may describe bremsstrahlung emission from relativistic electrons. The third model uses templates based on a detailed study of the gamma rays produced from the decay of pions originating from accelerated protons with an isotropic pitch angle distribution in a thick-target model~\citep[updated from][]{murp87}. We rely on the likelihood ratio test and the associated test statistic TS~\citep[]{Mattox:96} to estimate the significance of the detection. Here we define TS as twice the increment of the logarithm of the likelihood obtained by fitting the data with the source and background model component simultaneously. The TS of the power-law fit (TS$_{\rm PL}$) indicates the significance of the source detection under the assumption of a PL spectral shape and the $\Delta$TS=TS$_{\rm PLEXP}$-TS$_{\rm PL}$ quantifies how much more a complex spectral hypothesis improves the fit. Note that the significance ($\sigma$) can be roughly approximated as $\sqrt{\rm TS}$.

In Table~\ref{tab:SpectralAnalysis} we list the TS$_{\rm PL}$, $\Delta{\rm TS}$, $\Gamma$ the photon index for the best-fit model (PL when $\Delta{\rm TS}<25$ or PLEXP when $\Delta{\rm TS}\geq25$) and PLEXP cut-off energy. For several intervals  $\Delta{\rm TS}>$25, indicating that PLEXP provides a significantly better fit than PL. For these intervals we fit a series of pion-decay models to the data to determine the best proton spectral index following the same procedure described in~\citet{0004-637X-789-1-20}. Note that the TS values for PLEXP and pion-decay fits cannot be directly compared~\citep{wilks1938} because they are not nested models. However the PLEXP approximates the shape of the pion-decay spectrum; thus we expect the pion-decay models to provide a similarly acceptable fit. 

From both Table~\ref{tab:SpectralAnalysis} and Fig.~\ref{fig:SOL20170910_fullLC} and \ref{fig:SOL20170910_zoomLC} it is possible to note how the proton index softens during the impulsive phase of the flare (from 15:58-16:08 UT) plateaus from roughly 16:08 - 16:28 UT to a value of 4.0$\pm$0.1 and when the Sun come back into the field of view at 17:33 UT the proton index value is back to the initial value found during the impulsive phase and proceeds to soften once more. This behavior seems to imply there being two separate phases in the underlying acceleration agent of the protons.

\begin{deluxetable*}{ccccccc}[ht]
  \tablewidth{\textwidth}
  \tablecaption{\Fermi-LAT Spectral Analysis of the Solar flare of 2017 September 10}
  \tablehead{\colhead{Time Interval}  & \colhead{TS$_{PL}$} & \colhead{$\Delta$TS$^{a}$}  & \colhead{Photon Index$^{b}$} & \colhead{Cutoff Energy$^{c}$} & \colhead{Flux$^{d}$} & \colhead{Proton Index} \\
    \colhead{(UT)}  & \colhead{}  & \colhead{} & \colhead{}& \colhead{(MeV)} & \colhead{($\times$10$^{-5}$ ph cm$^{-2}$ s$^{-1}$)} & \colhead{}}
  \startdata
  \hline
15:56:55 -- 15:57:55 & 116 & 14 & -2.2$\pm$0.2 & - & 22$\pm$4 & - \\
15:57:55 -- 15:58:54 & 9600 & 118 & -0.8$\pm$0.2 & 211$\pm$30 & 463$\pm$18 & 3.8$\pm$0.1 \\
15:58:54 -- 15:59:54 & 38514 & 498 & -0.8$\pm$0.1 & 272$\pm$18 & 1306$\pm$27 & 3.2$\pm$0.1 \\
15:59:54 -- 16:00:54 & 42027 & 518 & -0.8$\pm$0.1 & 244$\pm$15 & 1319$\pm$26 & 3.5$\pm$0.1 \\
16:00:54 -- 16:01:53 & 26937 & 328 & -0.9$\pm$0.1 & 251$\pm$20 & 807$\pm$19 & 3.5$\pm$0.1 \\
16:01:53 -- 16:02:53 & 14323 & 256 & -0.6$\pm$0.1 & 194$\pm$18 & 477$\pm$14 & 3.6$\pm$0.1 \\
16:02:53 -- 16:04:52 & 3896 & 267 & -0.7$\pm$0.1 & 202$\pm$18 & 286$\pm$8 & 3.7$\pm$0.1 \\
16:04:52 -- 16:06:51 & 3225 & 212 & -0.7$\pm$0.1 & 194$\pm$19 & 194$\pm$6 & 3.9$\pm$0.1 \\
16:06:51 -- 16:08:50 & 3435 & 269 & -0.3$\pm$0.2 & 136$\pm$13 & 197$\pm$6 & 4.1$\pm$0.1 \\
16:08:50 -- 16:10:49 & 2864 & 241 & -0.4$\pm$0.2 & 147$\pm$14 & 169$\pm$5 & 4.1$\pm$0.1 \\
16:10:49 -- 16:12:48 & 3368 & 310 & -0.1$\pm$0.2 & 121$\pm$10 & 189$\pm$5 & 4.2$\pm$0.1 \\
16:12:48 -- 16:14:47 & 3136 & 231 & -0.6$\pm$0.2 & 156$\pm$15 & 191$\pm$6 & 4.2$\pm$0.1 \\
16:14:47 -- 16:16:46 & 3386 & 283 & -0.4$\pm$0.2 & 142$\pm$13 & 210$\pm$6 & 4.1$\pm$0.1 \\
16:16:46 -- 16:18:45 & 3091 & 283 & -0.3$\pm$0.2 & 135$\pm$12 & 215$\pm$6 & 4.1$\pm$0.1 \\
16:18:45 -- 16:20:44 & 2684 & 198 & -0.7$\pm$0.2 & 176$\pm$19 & 226$\pm$7 & 4.1$\pm$0.1 \\
16:20:44 -- 16:22:43 & 2223 & 217 & -0.2$\pm$0.2 & 136$\pm$14 & 231$\pm$8 & 4.1$\pm$0.1 \\
16:22:43 -- 16:24:42 & 1754 & 153 & -0.5$\pm$0.2 & 158$\pm$19 & 232$\pm$9 & 4.0$\pm$0.1 \\
16:24:42 -- 16:26:41 & 1254 & 94 & -0.7$\pm$0.2 & 185$\pm$28 & 266$\pm$13 & 4.0$\pm$0.2 \\
16:26:41 -- 16:28:40 & 871 & 58 & -0.8$\pm$0.3 & 197$\pm$37 & 338$\pm$21 & 4.0$\pm$0.2 \\
17:33:40 -- 17:58:16 & 6107 & 469 & -0.7$\pm$0.1 & 249$\pm$17 & 73$\pm$2 & 3.3$\pm$0.1 \\
19:03:16 -- 19:39:22 & 17051 & 1810 & -0.0$\pm$0.1 & 140$\pm$5 & 88$\pm$1 & 3.7$\pm$0.1 \\
20:44:22 -- 21:08:29 & 2309 & 277 & 0.1$\pm$0.2 & 117$\pm$11 & 35$\pm$1 & 4.2$\pm$0.1 \\
22:13:29 -- 22:49:35 & 2603 & 313 & 0.3$\pm$0.2 & 91$\pm$8 & 15.6$\pm$0.6 & 4.7$\pm$0.2 \\
23:54:47 -- 00:18:47 & 311 & 68 & 2.0$\pm$0.9 & 55$\pm$11 & 5.6$\pm$0.6 & 4.9$\pm$0.4 \\
01:23:51 -- 02:00:21$^{e}$ & 283 & 55 & 1.7$\pm$0.8 & 48$\pm$10 & 2.5$\pm$0.2 & 6.0$\pm$0.5 \\
03:05:44 -- 03:29:14$^{e}$ & 59 & 12 & -2.6$\pm$0.2 & - & 1.1$\pm$0.3 & - \\
04:34:04 -- 05:11:04$^{e}$ & 39 & 6 & -2.7$\pm$0.2 & -  & 0.5$\pm$0.1 & -  \\
  \enddata
\label{tab:SpectralAnalysis}
\tablenotetext{a}{$\Delta$TS=TS$_{\rm PLEXP}$-TS$_{\rm PL}$}
\tablenotetext{b}{Photon index from best-fit model. The PL is defined as $\frac{dN(E)}{dE} = N_0E^{\Gamma}$ and the PLEXP as $\frac{dN(E)}{dE} = N_0E^{\Gamma}\exp(-\frac{E}{E_{c}})$ where $E_c$ is the cutoff energy.}
\tablenotetext{c}{From the fit with the PLEXP model.}
\tablenotetext{d}{Integrated flux between 100\,MeV and 10\,GeV calculated for the best-fit model.}
\tablenotetext{e}{These intervals are during 2017 September 11.}
\end{deluxetable*}

\subsection{Localization of the emission}

This flare is bright enough to do a time resolved localization study. Similarly to the situation of the 2012 March 7 flares, we face the same difficulties with the fisheye correction: the asymmetric observational profile induces uncertainties that change the apparent position of the gamma-ray source at each time windows, but in a different direction from one orbit to the next~\citep{2012ApJS..203....4A}. We performed a study on the impact of the fisheye effect on the position of the gamma-ray source as a function of the minimum energy threshold used. We find that the uncorrected position changes significantly as the energy threshold is increased, which is what we expect from the fisheye effect: the correction at 60 MeV changes the position by more than two 68\% error radii whereas the correction at 300 MeV remains within the 68\% error radius. When examining the corrected positions we find that they are somewhat overlapping, with the position of the gamma-ray source above 300 MeV. This could indicate that the systematic error due to the fisheye effect is larger than the 68\% statistical error, for this reason we will therefore use the 95\% error radius. 

\begin{figure*}[ht!]   
\begin{center}   
\includegraphics[width=6 in]{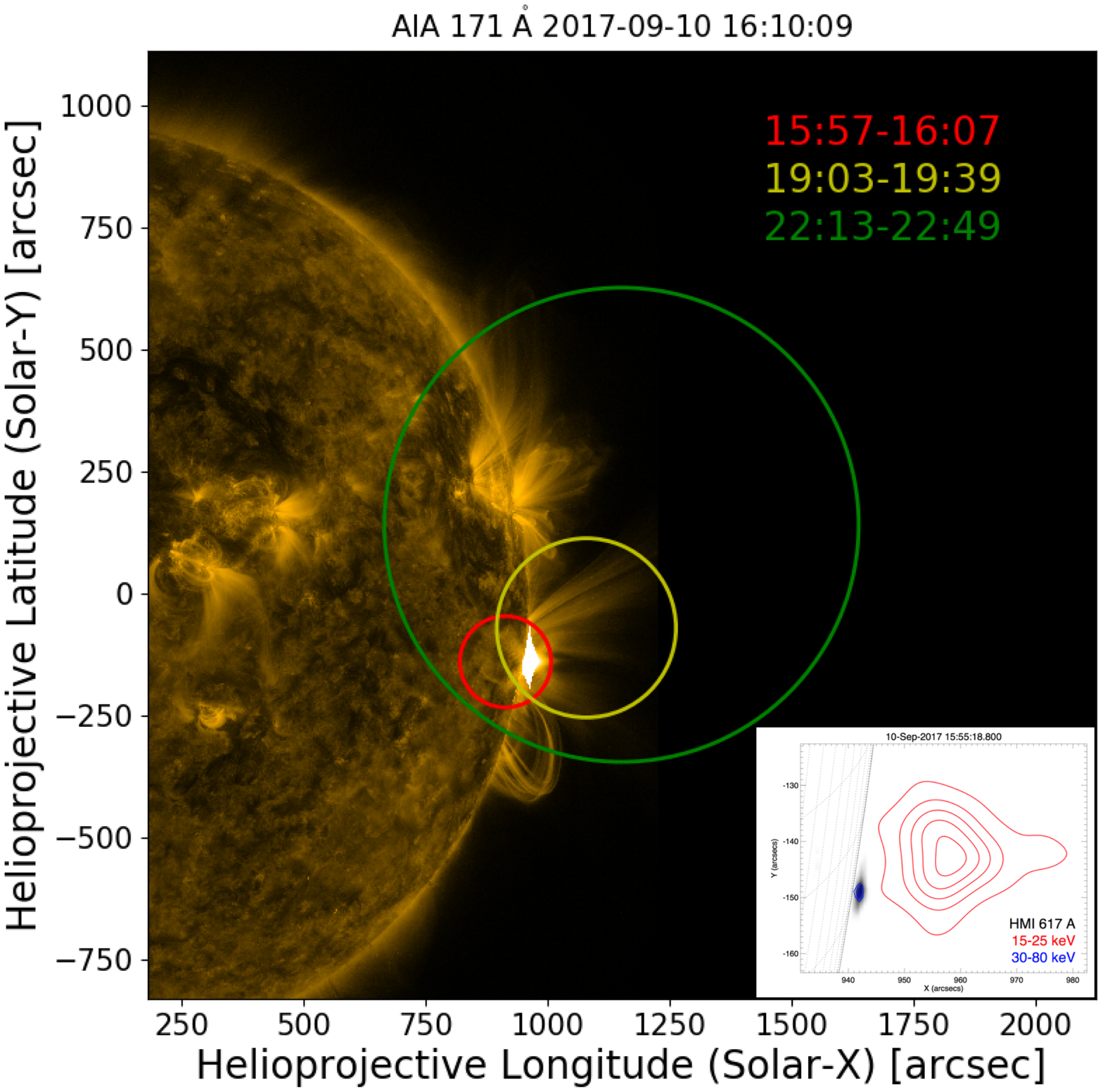}  
\caption{Localization evolution for three different time intervals, overlaid on the SDO AIA 171 image of the Sun at 16:10:09UT. 
On the lower right corner: \hsi contours in the thermal (red) and non-thermal (blue) range are shown on a HMI difference image (dark is enhanced emissions). The scale of the insert is such that both the thermal and non-thermal emissions are contained in the LAT 95\% error circle.}
\label{sept10_spatial_evolution}   
\end{center}   
\end{figure*}

In addition, the Fermi-LAT localization capabilities are limited by some small systematics errors due to the instrument and the spacecraft alignment precision , which consist in a 1.1 scale factor on the localization error, and an additional 10$\arcsec$ on the error radius~\citep{2012ApJS..199...31N}. These factors are added in quadrature to the 95\% error radii shown in Figure~\ref{sept10_spatial_evolution}.

\begin{deluxetable*}{lccccc}[ht!]
  \tablewidth{\textwidth}
  \tablecaption{Evolution of the gamma ray source localization with fisheye correction for the 2017-09-10 flare}
  \tablehead{\colhead{Date and Time}  & \colhead{HelioX} & \colhead{HelioY}  & \colhead{Localization Error (95\% c.l.)} & \colhead{Distance from AR} & \colhead{Relative Distance} \\
    \colhead{2017-09-10 (UT)}  & \colhead{arc sec}  & \colhead{arc sec} & \colhead{arc sec}& \colhead{arc sec} & \colhead{}}
  \startdata
  \hline
   15:56:55 -- 16:06:51 &    910 &    - 140 &  90 &     80 &                    0.8  \\
   19:03:16 -- 19:39:22 &    1090 &     -70 &  180 &    150 &                 0.8  \\
   22:13:29 -- 22:49:35 &    1150 &     140 &  490 &    330 &                 0.7 \\
  \enddata
  \label{tab:sept10_localization}
\end{deluxetable*}

We limit our localization study to the time intervals with longer exposure ($>$30 minutes) and smaller average boresight angles ($<$55$^{\circ}$). We select the time windows starting at 15:56UT, 19:03UT and 22:13UT and use an energy threshold of 300 MeV.
For the first time window we compute the position using the S15 event class immune to pile-up effects and maximize the number of photons collected.

 Table~\ref{tab:sept10_localization} gives the corrected positions of the gamma-ray emission for those times\footnote{We list the mean time of the interval in the table and in Figure~\ref{sept10_spatial_evolution}.} and the 95\% containment radius (statistical only). In the table we also give the distance between the position of the AR (estimated to (HelioX,HelioY=957,-135 arc seconds) and the best position for the gamma-ray source. The last column shows the ratio between this distance and the 95\% containment radius. We see that the location of the gamma-ray emission is consistent with the AR for all time windows, which is different than what observed for the 2012 March 7 \citep{0004-637X-789-1-20}.

\section{Discussion}
The 2017 September 10 solar flare was an exceptional flare. It was the brightest gamma-ray source in the sky for more than 12 hours and it was associated with the second GLE of this solar cycle (GLE 72). The Sun was in the field of view of the LAT for both the impulsive and extended phase of the solar flare allowing for very good coverage of the event. The flux behavior of this flare is similar to other long duration flares detected by the LAT, namely there is sharp rising and descending peak coincident with the X-ray flaring activity followed by a slow rise and fall in flux over a period of several hours. However, the temporal variation of the estimated proton index does not show a continuous softening with time as was seen for other long duration flares \citep[such the 2012 March 7 flares,][]{0004-637X-789-1-20} but instead there appears to be three phases in the evolution of the proton index. Between 15:48 - 16:08 UT, in coincidence with the descending phase of the hard X-ray flare time profile, the proton index softens with values ranging from 3.2$\pm$0.1 to 4.0$\pm$0.1. During the twenty minutes that follow, both the proton index and the gamma-ray flux appear to harden/increase and when the Sun comes back into the field of view of the LAT at 17:33 UT, the proton index has once again hardened to a value of 3.3$\pm$0.1 and proceeds to soften for the remaining almost 9 hours of gamma-ray flaring activity. 
This behavior in the proton index variations seems to suggest two different protons acceleration mechanisms.

Based on the analysis performed by Mishev et al. (paper submitted to Solar Physics) we know that the onset time of GLE 72 as measured by 27 neutron monitors distributed over the globe starts at 16:30 UT $\pm$1 minute (these are proton arrival times at 1 AU so that the arrival time is delayed by roughly 3 minutes with respect the gamma-ray photons).  In their analysis of GLE 72, Mishev et al. find that the ratio of Fe/O of the SEPs is low ($<$0.07 in the 50-100 MeV/n range) consistent with a gradual event and the timing of the GLE 72 is also consistent with a hypothesis of particle acceleration at a coronal shock driven by the CME. Unfortunately there is a gap in the LAT data from 16:28 - 17:33 UT because the Sun was not in the field of view. However, based simply on the timing of the events, the third phase of the proton index variation occurs after the GLE onset as well as after the first LASCO C2 appearance of the CME associated with this flare. During this time the X-ray flare has ceased and no other flaring activity is visible. If indeed the protons responsible for the LDGRF emission share the same acceleration mechanism as the SEP population measured at 1 AU then this could also explain the detection of the long duration high-energy emission observed in this flare.

Another unique aspect of this flare is that the location of the gamma-ray emission is consistent with the AR over a time period of more than 6 hours as can be seen in Fig.~\ref{sept10_spatial_evolution}. This behavior is different from what was found for the 2012 March 7 flares~\citep{0004-637X-789-1-20} where the emission position was found to move across the solar disk with time. This most likely an effect due to the position of the AR on the western limb of the Sun, and that the LAT can only detect the gamma-rays originating from the protons interacting in the chromosphere on the visible side of the Sun.

The \hsi observations clearly show that the detected non-thermal HXR emissions from the flare ribbons occur above the limb as seen from Earth, without emission from the solar disk (see insert at bottom right in Figure~\ref{sept10_spatial_evolution}). As gamma-ray producing ions are typically stopped at much lower altitudes than hard X-ray producing electrons, it is expected that gamma-ray photons from the flare ribbon traveling towards Earth are heavily attenuated or they may even be completely stopped before they escape towards Earth. It is therefore questionable that the initial impulsive peak seen by LAT is from the flare ribbons. The time delay between the 50 keV hard X-rays and the LAT peak of about 1 minute (see Figure~\ref{fig:SOL20170910_zoomLC}) can be seen as a further hint that the 100 MeV producing protons could be from a different population than the flare-accelerated particles that precipitate towards the flare ribbons. However, the currently available data is inconclusive regarding the nature of the initial impulsive component.


A more detailed analysis of the gamma-ray flare together with the GLE and CME properties is necessary in order to better understand the connection between the accelerating agents of the two proton populations responsible for the emission detected by the LAT and the SEP measured at Earth. This is the first time that a long duration gamma-ray flare has been detected together with a GLE and serves as a precious case study for the acceleration mechanisms at work.

\acknowledgements
The \Fermi-LAT Collaboration acknowledges support for LAT development, operation and data analysis from NASA and DOE (United States), CEA/Irfu and IN2P3/CNRS (France), ASI and INFN (Italy), MEXT, KEK, and JAXA (Japan), and the K.A.~Wallenberg Foundation, the Swedish Research Council and the National Space Board (Sweden). Science analysis support in the operations phase from INAF (Italy) and CNES (France) is also gratefully acknowledged. This work performed in part under DOE Contract DE-AC02-76SF00515.

%
\bibliography{SOL170910}

\newcommand{\noop}[1]{}
\begin{thebibliography}{15}
\expandafter\ifx\csname natexlab\endcsname\relax\def\natexlab#1{#1}\fi

\bibitem[{{Ackermann} {et~al.}(2012){Ackermann}, {Ajello}, {Albert},
  {Allafort}, {Atwood}, {Axelsson}, {Baldini}, {Ballet}, {Barbiellini},
  {Bastieri}, {Bechtol}, {Bellazzini}, {Bissaldi}, {Blandford}, {Bloom},
  {Bogart}, {Bonamente}, {Borgland}, {Bottacini}, {Bouvier}, {Brandt},
  {Bregeon}, {Brigida}, {Bruel}, {Buehler}, {Burnett}, {Buson}, {Caliandro},
  {Cameron}, {Caraveo}, {Casandjian}, {Cavazzuti}, {Cecchi}, {{\c C}elik},
  {Charles}, {Chaves}, {Chekhtman}, {Cheung}, {Chiang}, {Ciprini}, {Claus},
  {Cohen-Tanugi}, {Conrad}, {Corbet}, {Cutini}, {D'Ammando}, {Davis}, {de
  Angelis}, {DeKlotz}, {de Palma}, {Dermer}, {Digel}, {Silva}, {Drell},
  {Drlica-Wagner}, {Dubois}, {Favuzzi}, {Fegan}, {Ferrara}, {Focke}, {Fortin},
  {Fukazawa}, {Funk}, {Fusco}, {Gargano}, {Gasparrini}, {Gehrels}, {Giebels},
  {Giglietto}, {Giordano}, {Giroletti}, {Glanzman}, {Godfrey}, {Grenier},
  {Grove}, {Guiriec}, {Hadasch}, {Hayashida}, {Hays}, {Horan}, {Hou}, {Hughes},
  {Jackson}, {Jogler}, {J{\'o}hannesson}, {Johnson}, {Johnson}, {Johnson},
  {Kamae}, {Katagiri}, {Kataoka}, {Kerr}, {Kn{\"o}dlseder}, {Kuss}, {Lande},
  {Larsson}, {Latronico}, {Lavalley}, {Lemoine-Goumard}, {Longo}, {Loparco},
  {Lott}, {Lovellette}, {Lubrano}, {Mazziotta}, {McConville}, {McEnery},
  {Mehault}, {Michelson}, {Mitthumsiri}, {Mizuno}, {Moiseev}, {Monte},
  {Monzani}, {Morselli}, {Moskalenko}, {Murgia}, {Naumann-Godo}, {Nemmen},
  {Nishino}, {Norris}, {Nuss}, {Ohno}, {Ohsugi}, {Okumura}, {Omodei},
  {Orienti}, {Orlando}, {Ormes}, {Paneque}, {Panetta}, {Perkins},
  {Pesce-Rollins}, {Pierbattista}, {Piron}, {Pivato}, {Porter}, {Racusin},
  {Rain{\`o}}, {Rando}, {Razzano}, {Razzaque}, {Reimer}, {Reimer}, {Reposeur},
  {Reyes}, {Ritz}, {Rochester}, {Romoli}, {Roth}, {Sadrozinski}, {Sanchez},
  {Saz Parkinson}, {Sbarra}, {Scargle}, {Sgr{\`o}}, {Siegal-Gaskins},
  {Siskind}, {Spandre}, {Spinelli}, {Stephens}, {Suson}, {Tajima}, {Takahashi},
  {Tanaka}, {Thayer}, {Thayer}, {Thompson}, {Tibaldo}, {Tinivella}, {Tosti},
  {Troja}, {Usher}, {Vandenbroucke}, {Van Klaveren}, {Vasileiou}, {Vianello},
  {Vitale}, {Waite}, {Wallace}, {Winer}, {Wood}, {Wood}, {Wood}, {Yang}, \&
  {Zimmer}}]{2012ApJS..203....4A}
{Ackermann}, M., {Ajello}, M., {Albert}, A., {et~al.} 2012, \apjs, 203, 4

\bibitem[{{Ackermann} {et~al.}(2014){Ackermann}, Ajello, Albert, Allafort,
  Baldini, Barbiellini, Bastieri, Bechtol, Bellazzini, Bissaldi, Bonamente,
  Bottacini, Bouvier, Brandt, Bregeon, Brigida, Bruel, Buehler, Buson,
  Caliandro, Cameron, Caraveo, Cecchi, Charles, Chekhtman, Chen, Chiang,
  Chiaro, Ciprini, Claus, Cohen-Tanugi, Conrad, Cutini, D'Ammando, de~Angelis,
  de~Palma, Dermer, Desiante, Digel, Venere, do~Couto~e Silva, Drell,
  Drlica-Wagner, Favuzzi, Fegan, Focke, Franckowiak, Fukazawa, Funk, Fusco,
  Gargano, Gasparrini, Germani, Giglietto, Giordano, Giroletti, Glanzman,
  Godfrey, Grenier, Grove, Guiriec, Hadasch, Hayashida, Hays, Horan, Hughes,
  Inoue, Jackson, Jogler, J{\'o}hannesson, Johnson, Kamae, Kawano,
  Kn{\"o}dlseder, Kuss, Lande, Larsson, Latronico, Lemoine-Goumard, Longo,
  Loparco, Lott, Lovellette, Lubrano, Mayer, Mazziotta, McEnery, Michelson,
  Mizuno, Moiseev, Monte, Monzani, Moretti, Morselli, Moskalenko, Murgia,
  Murphy, Nemmen, Nuss, Ohno, Ohsugi, Okumura, Omodei, Orienti, Orlando, Ormes,
  Paneque, Panetta, Perkins, Pesce-Rollins, Petrosian, Piron, Pivato, Porter,
  Rain{\`o}, Rando, Razzano, Reimer, Reimer, Ritz, Schulz, Sgr{\`o}, Siskind,
  Spandre, Spinelli, Takahashi, Takeuchi, Tanaka, Thayer, Thayer, Thompson,
  Tibaldo, Tinivella, Tosti, Troja, Tronconi, Usher, Vandenbroucke, Vasileiou,
  Vianello, Vitale, Werner, Winer, Wood, Wood, Wood, \&
  Yang}]{0004-637X-787-1-15}
{Ackermann}, M., Ajello, M., Albert, A., {et~al.} 2014, \apj, 787, 15

\bibitem[{{Ackermann} {et~al.}(2017){Ackermann}, {Allafort}, {Baldini},
  {Barbiellini}, {Bastieri}, {Bellazzini}, {Bissaldi}, {Bonino}, {Bottacini},
  {Bregeon}, {Bruel}, {Buehler}, {Cameron}, {Caragiulo}, {Caraveo},
  {Cavazzuti}, {Cecchi}, {Charles}, {Ciprini}, {Costanza}, {Cutini},
  {D'Ammando}, {de Palma}, {Desiante}, {Digel}, {Di Lalla}, {Di Mauro}, {Di
  Venere}, {Drell}, {Favuzzi}, {Fukazawa}, {Fusco}, {Gargano}, {Giglietto},
  {Giordano}, {Giroletti}, {Grenier}, {Guillemot}, {Guiriec}, {Jogler},
  {J{\'o}hannesson}, {Kashapova}, {Krucker}, {Kuss}, {La Mura}, {Larsson},
  {Latronico}, {Li}, {Liu}, {Longo}, {Loparco}, {Lubrano}, {Magill}, {Maldera},
  {Manfreda}, {Mazziotta}, {Mitthumsiri}, {Mizuno}, {Monzani}, {Morselli},
  {Moskalenko}, {Negro}, {Nuss}, {Ohsugi}, {Omodei}, {Orlando}, {Pal'shin},
  {Paneque}, {Perkins}, {Pesce-Rollins}, {Petrosian}, {Piron}, {Principe},
  {Rain{\`o}}, {Rando}, {Razzano}, {Reimer}, {Rubio da Costa}, {Sgr{\`o}},
  {Simone}, {Siskind}, {Spada}, {Spandre}, {Spinelli}, {Tajima}, {Thayer},
  {Torres}, {Troja}, \& {Vianello}}]{2017ApJ...835..219A}
{Ackermann}, M., {Allafort}, A., {Baldini}, L., {et~al.} 2017, \apj, 835, 219

\bibitem[{{Ajello} {et~al.}(2014){Ajello}, Allafort, Baldini, Barbiellini,
  Bastieri, Bellazzini, Bissaldi, Bonamente, Brandt, Bregeon, Brigida, Bruel,
  Buehler, Buson, Caliandro, Cameron, Caraveo, Cecchi, Charles, Chekhtman,
  Chiang, Chiaro, Ciprini, Claus, Cohen-Tanugi, Cominsky, Conrad, Cutini,
  D'Ammando, de~Palma, Dermer, Desiante, Digel, do~Couto~e Silva, Drell,
  Drlica-Wagner, Favuzzi, Focke, Franckowiak, Fukazawa, Fusco, Gargano,
  Gasparrini, Germani, Giglietto, Giommi, Giordano, Giroletti, Glanzman,
  Godfrey, Grenier, Grove, Guiriec, Hadasch, Hayashida, Hays, Horan, Hou,
  Hughes, Inoue, Jackson, Jogler, J{\'o}hannesson, Johnson, Johnson, Kamae,
  Kn{\"o}dlseder, Kocevski, Kuss, Lande, Larsson, Latronico, Longo, Loparco,
  Lott, Lovellette, Lubrano, Mayer, Mazziotta, McEnery, Michelson, Mizuno,
  Moiseev, Monte, Monzani, Morselli, Moskalenko, Murgia, Murphy, Nakamori,
  Nemmen, Nuss, Ohno, Ohsugi, Omodei, Orienti, Orlando, Ormes, Paneque,
  Panetta, Perkins, Pesce-Rollins, Petrosian, Piron, Pivato, Porter, Rain{\`o},
  Rando, Razzano, Reimer, Reimer, Roth, Schulz, Sgr{\`o}, Siskind, Spandre,
  Spinelli, Takahashi, Thayer, Thayer, Thompson, Tibaldo, Tinivella, Tosti,
  Troja, Usher, Vandenbroucke, Vasileiou, Vianello, Vitale, Werner, Winer,
  Wood, Wood, \& Yang}]{0004-637X-789-1-20}
{Ajello}, M. A.~A., Allafort, A., Baldini, L., {et~al.} 2014, \apj, 789, 20

\bibitem[{{Atwood} {et~al.}(2009){Atwood}, {Ackermann}, {Ajello}, {Baldini},
  {Ballet}, {Barbiellini}, {Baring}, {Bastieri}, {Bechtol}, {Bellazzini},
  {Berenji}, {Bhat}, {Bissaldi}, {Blandford}, {Bonamente}, {Bonnell},
  {Borgland}, {Bouvier}, {Bregeon}, {Brigida}, {Bruel}, {Buehler}, {Buson},
  {Caliandro}, {Cameron}, {Caraveo}, {Casandjian}, {Cecchi}, {Charles},
  {Chekhtman}, {Chiang}, {Ciprini}, {Claus}, {Connaughton}, {Conrad}, {Cutini},
  {de Angelis}, {de Palma}, {Dermer}, {Silva}, {Drell}, {Dubois}, {Favuzzi},
  {Fukazawa}, {Fusco}, {Gargano}, {Gehrels}, {Germani}, {Giglietto}, {Giommi},
  {Giordano}, {Giroletti}, {Glanzman}, { Godfrey}, {Granot}, {Grenier},
  {Guiriec}, {Hadasch}, {Hanabata}, {Hughes}, {J{\'o}hannesson}, {Johnson},
  {Kamae}, {Katagiri}, {Kataoka}, {Kerr}, {Kn{\"o}dlseder}, {Kuss}, {Lande},
  {Latronico}, {Lee}, {Longo}, {Loparco}, {Lott}, {Lubrano}, {Mazziotta},
  {McEnery}, {M{\'e}sz{\'a}ros}, {Michelson}, {Mizuno}, {Moiseev}, {Monzani},
  {Morselli}, {Moskalenko}, {Murgia}, {Nakamori}, {Naumann-Godo}, {Nolan},
  {Norris}, {Nuss}, {Ohsugi}, {Okumura}, {Omodei}, {Orlando}, {Paciesas},
  {Pelassa}, {Pesce-Rollins}, {Pierbattista}, {Piron}, {Porter}, {Racusin},
  {Rain{\`o}}, {Razzano}, {Razzaque}, {Reimer}, {Reimer}, { Reyes}, {Roth},
  {Sadrozinski}, {Sgr{\`o}}, {Siskind}, {Smith}, {Sonbas}, {Spandre},
  {Spinelli}, {Stamatikos}, {Strickman}, {Takahashi}, {Tanaka}, {Tanaka},
  {Thayer}, {Thayer}, {Torres}, {Tosti}, {Troja}, {Uehara}, {Usher},
  {Vandenbroucke}, {Vasileiou}, {Vianello}, {Vilchez}, {Vitale}, {von Kienlin},
  {Waite}, {Wang}, {Winer}, {Wood}, {Yamazaki}, {Yang}, {Ziegler}, {Piro}, \&
  {Fermi Collaboration}}]{LATPaper}
{Atwood}, W.~B.{Abdo}, A.~A., {Ackermann}, M., {Ajello}, M., {et~al.} 2009,
  \apj, 697, 1071

\bibitem[{{Chupp} \& {Ryan}(2009)}]{chup09}
{Chupp}, E.~L., \& {Ryan}, J.~M. 2009, Research in Astronomy and Astrophysics,
  9, 11

\bibitem[{{Forrest} {et~al.}(1985){Forrest}, {Vestrand}, {Chupp}, {Rieger},
  {Cooper}, \& {Share}}]{forr85}
{Forrest}, D.~J., {Vestrand}, W.~T., {Chupp}, E.~L., {et~al.} 1985, in
  International Cosmic Ray Conference, Vol.~4, International Cosmic Ray
  Conference, ed. {M.~Garcia-Munoz, K.~R.~Pyle, \& J.~A.~Simpson}, 146--149

\bibitem[{{Kanbach} {et~al.}(1993){Kanbach}, {Bertsch}, {Fichtel}, {Hartman},
  {Hunter}, {Kniffen}, {Kwok}, {Lin}, {Mattox}, \&
  {Mayer-Hasselwander}}]{1993A&AS...97..349K}
{Kanbach}, G., {Bertsch}, D.~L., {Fichtel}, C.~E., {et~al.} 1993, \aaps, 97,
  349

\bibitem[{{Krucker} {et~al.}(2015){Krucker}, {Saint-Hilaire}, {Hudson},
  {Haberreiter}, {Martinez-Oliveros}, {Fivian}, {Hurford}, {Kleint},
  {Battaglia}, {Kuhar}, \& {Arnold}}]{2015ApJ...802...19K}
{Krucker}, S., {Saint-Hilaire}, P., {Hudson}, H.~S., {et~al.} 2015, \apj, 802,
  19

\bibitem[{Mattox {et~al.}(1996)Mattox, Bertsch, Chiang, Dingus, Digel,
  Esposito, Fierro, Hartman, Hunter, Kanbach, Kniffen, Lin, Macomb,
  Mayer-Hasselwander, Michelson, von Montigny, Mukherjee, Nolan, Ramanamurthy,
  Schneid, Sreekumar, Thompson, \& Willis}]{Mattox:96}
Mattox, J.~R., Bertsch, D.~L., Chiang, J., {et~al.} 1996, \apj, 461, 396

\bibitem[{{Murphy} {et~al.}(1987){Murphy}, {Dermer}, \& {Ramaty}}]{murp87}
{Murphy}, R.~J., {Dermer}, C.~D., \& {Ramaty}, R. 1987, \apjs, 63, 721

\bibitem[{{Nolan} {et~al.}(2012){Nolan}, {Abdo}, {Ackermann}, {Ajello},
  {Allafort}, {Antolini}, {Atwood}, {Axelsson}, {Baldini}, {Ballet}, \&
  et~al.}]{2012ApJS..199...31N}
{Nolan}, P.~L., {Abdo}, A.~A., {Ackermann}, M., {et~al.} 2012, \apjs, 199, 31

\bibitem[{{Ramaty} {et~al.}(1987){Ramaty}, {Murphy}, \& {Dermer}}]{ram87}
{Ramaty}, R., {Murphy}, R.~J., \& {Dermer}, C.~D. 1987, \apjl, 316, L41

\bibitem[{{Ryan}(2000)}]{2000SSRv...93..581R}
{Ryan}, J.~M. 2000, \ssr, 93, 581

\bibitem[{{Wilks}(1938)}]{wilks1938}
{Wilks}, S.~S. 1938, Ann. Math. Stat., 9, 60

\end{thebibliography}
\bibliographystyle{apj}

\end{document}